\documentclass[11pt,dvips]{article}
\usepackage{amssymb}
\usepackage{amsmath}
\usepackage{graphicx}
\usepackage{bm}
\newcommand{\mathsym}[1]{{}}

\usepackage{graphicx}
\usepackage{amscd}
\usepackage{rotating}
\usepackage{color}

\newcommand{\bra}{\begin{array}}
\newcommand{\era}{\end{array}}
\newcommand{\beq}{\begin{equation}}
\newcommand{\eeq}{\end{equation}}
\newcommand{\beqar}{\begin{eqnarray}}
\newcommand{\eeqar}{\end{eqnarray}}

\def\BC{\bb C}
\def\_\BC{\bbi C}

\def\( {\left(}
   \def\) {\right)}
\def\[ {\left[}
\def\] {\right]}

\def\ep {{\epsilon}}

\def \A {{\mathcal A}}

\newcommand{\si}{\sigma}

\newcommand{\be}{\beta}

\newcommand{\ga}{\gamma}
\newcommand{\te}{\theta}

\newcommand{\pa}{\partial}
\newcommand{\al}{\alpha}

\newcommand{\del}{\delta}

\newcommand{\ti}{\tilde}
\newcommand{\da}{\dagger}

\newcommand{\lb}{\label}
\newcommand{\h}{\hat}
\newcommand{\fr}{\frac}

\begin{document}
\begin{titlepage}
\setcounter{page}{1}
\renewcommand{\thefootnote}{\fnsymbol{footnote}}

\vspace{5mm}

\begin{center}

{\Large\bf A Noncommutative Space Approach to Confined \\ Dirac Fermions in Graphene }

\vspace{0.5cm}

{\bf{\"{O}mer F. Dayi}$^{a,b}$\footnote{{\textsf dayi@itu.edu.tr and dayi@gursey.gov.tr}}}
\,{and}\,
{ \bf{Ahmed Jellal}$^{c,d}$\footnote{{\textsf jellal@pks.mpg.de and jellal@ucd.ma}}}
\vspace{5mm}

{\em $^{a}${\it Physics Department, Faculty of Science and
Letters, Istanbul Technical University,\\
TR-34469 Maslak--Istanbul, Turkey} }

{\em $^{b}${\it Feza G\"{u}rsey Institute, P.O. Box 6, TR-34349,
\c{C}engelk\"{o}y, Istanbul, Turkey } }

{\em $^{c}${\it Physics Department, College of Sciences, King Faisal University,\\
PO Box 380, Alahsa 31982,
Saudi Arabia}}

{\em $^{d}${\it Theoretical Physics Group,
Faculty of Sciences, Choua\"ib Doukkali University,\\ 
PO Box 20,
24000 El Jadida,
Morocco}}

\vspace{3cm}

\begin{abstract}
A  generalized algebra  of noncommutative  coordinates and momenta embracing  non-Abelian gauge fields  is  proposed.
Through  a two-dimensional realization of this algebra  for a gauge field
including electromagnetic vector potential and  two spin-orbit-like coupling terms,
 a  Dirac-like Hamiltonian in noncommutative coordinates is introduced.
We established the corresponding energy spectrum
and from that we derived
the relation between the energy level quantum number and  the magnetic field at the maxima of  Shubnikov-de Haas oscillations.
By tuning  the noncommutativity parameter $\theta$ in terms of the values of
magnetic field at the maxima of Shubnikov-de Haas oscillations, we accomplished
the experimentally observed Landau plot of the peaks for graphene. Accepting that the experimentally observed behavior
is due to the confinement of carriers, we  conclude that our method of introducing
noncommutative coordinates provides another
formulation of the confined massless Dirac fermions  in graphene.

\end{abstract}
\end{center}
\end{titlepage}

\newpage


\section{Introduction}

The recent experimental observations
of the anomalous quantum Hall effect in monocrystalline graphite films of one atomic layer  thickness~\cite{novoselov,zhang}
revealed the fact that  in this material, called graphene, electrons behave as effectively
massless  relativistic particles.
 Theoretically, this unexpected quantization of Hall conductivity
can be explained in terms of the massless Dirac-like theory~\cite{ando,sharapov}.
On the other hand magnetic oscillations of electrical conductivity known as Shubnikov-de Haas (SdH) oscillations~\cite{haas}
were measured in patterned
epitaxial graphene~\cite{berger}. It was shown that its transport properties
result from carrier confinement and coherence.
Moreover, in~\cite{berger}, to explain the observed behavior of the maxima of SdH oscillations,
an analytic expression for the energy levels which takes into account the confinement of
the charge carriers due to the micrometer-scale of the sample, has been proposed.
In fact, it is in accord with the theoretical study of
confining massless Dirac fermions by introducing a coordinate dependent mass term~\cite{peres}.

Noncommutativity of coordinates  naturally  appear
in Landau problem as well as in string theories where  noncommutativity
is proportional to the magnetic field in the former and  to the background fields in the latter. However, one can  also  introduce noncommutativity of
coordinates by the constant noncommutativity parameter $\theta $ as an intrinsic property of the space.  Then  it is legitimate to consider
noncommutative version of any physical system employing appropriate formulation. When noncommutativity of space is imposed without
attributing a definite physical meaning to the noncommutativity parameter $\theta$  one should provide its interpretation. Obviously, noncommutativity  can be taken as a fundamental property
of space by claiming that  the noncommutativity parameter $\theta $ is very small and it is responsible for the errors in the measurements
of the related physical quantities.  In this case each physical system will provide a different limit for the possible value of
the noncommutativity parameter $\theta $ (see e.g. \cite{o3} and references therein). However, there exists another interpretation:
Noncommutative formulation of a dynamical system can be considered as a tool to link some diverse manifestations of a physical phenomena, e.g.
obtaining the fractional Hall effect from
the Hall effect in noncommutative coordinates
by an appropriate choice of the  noncommutativity parameter $\theta ,$ as was reported in \cite{ao}.
Because of retaining the terms up to a definite order in $\theta$ one should show that in the dynamical problem considered
the fixed value of $\theta$ is in accord with this approximation.
This interpretation of noncommutating coordinates as a linkage between different phenomena may give same clues of finding an easier
method of formulating interacting systems from noninteracting theories~\cite{sc}.
Hence, we would like to study  massless Dirac theory in two-dimensional noncommutative space
to perceive whether a similar interpretation of noncommutativity is possible which can yield a better understanding of
some peculiar properties of graphene.
However, introducing
noncommutative coordinates into
spin-dependent Hamiltonian systems defined by constant matrices is not well established.
The usual method of introducing noncommutativity is to  replace ordinary products with
star products which is equivalent to the shift
\beq
\lb{bsh}
 x_\mu\rightarrow  x_\mu-\frac{1}{2\hbar}\theta_{\mu \nu}   p^\nu ,
\eeq
where $(x_\mu , p_\mu )$ are
the  quantum phase space variables.
Obviously, this method, which does not take into account spin degrees of freedom, is not suitable to
deal with  matrix valued, constant  observables.

 Utilizing the semiclassical techniques developed in~\cite{o1},
a method of introducing noncommutativity of space  appropriate to deal with
non-Abelian  spin matrices was presented in~\cite{oe,o3}. However, this semiclassical treatment is not
amenable to introduce noncommutativity
into Dirac equation of fermions interacting with external fields.
We propose a solution for this issue.
First   quantum commutation relations between   phase space variables in noncommutative space
suitable to consider non-Abelian  fields are presented. We derived them from the semiclassical brackets of
classical phase space  coordinates  proposed in~\cite{oe}. Then, realizations of this algebra
may be employed to introduce noncommutative coordinates in Hamiltonian systems.
It is applicable to systems where the interaction terms are  coordinate independent  but  non-Abelian
because of being matrices. Obviously, this formalism can be applied to Hamiltonian systems whose interaction terms are not matrices, so that
it establishes an alternative to the custom star product approach which is equivalent to the  shift (\ref{bsh}).
We then focus on the confinement problem
of massless Dirac particles in graphene
and propose that introducing noncommutative coordinates within our approach
may be used to deal with one of its basic features: the Landau plot of the maxima of SdH oscillations. To obtain the $\theta$ deformation
we give a realization of the  generalized  canonical commutation relations
in the presence of spin-orbit-like couplings and a transverse magnetic field.
We obtain the spectrum of the proposed Hamiltonian and show that it  can be used to formulate
the SdH effect in graphene.
By fixing the noncommutativity parameter $\theta$ we actually established  a good agreement
with the experimental observations which are known to result from
the confinement of massless Dirac fermions.
In the last section we discuss the obtained results and their future applications.

\section{Generalized algebra}

We would like to present  a formulation of quantum mechanics
in noncommutative coordinates acquired by quantizing the semiclassical approach of \cite{o1}.
Hence, let us briefly recall this formulation. When one considers a dynamical system in   the classical
phase space  $(P_I , Q_I );\ I =1,\cdots ,M$ one can introduce the star product
\begin{equation}
\label{SCT}
\star   =
\exp \left[
\frac{i\hbar }{2} \left(
\frac{\overleftarrow{\partial}}{\partial Q^I}
\frac{\overrightarrow{\partial}}{\partial P_I}
-\frac{\overleftarrow{\partial}}{\partial P_I}
\frac{\overrightarrow{\partial}}{\partial Q^I}
\right) \right] ,
\end{equation}
to achieve quantization of the system within the Weyl-Wigner-Groenewold-Moyal approach~\cite{WWGM}.
Corresponding to the quantum commutators one considers the Moyal bracket of the
observables $f(P,Q )$ and $g(P,Q )$  which are some functions as
$$
\left[f(P,Q ) , g(P,Q ) \right]_\star \equiv f(P,Q) \star g(P,Q ) - g(P,Q ) \star f(P,Q ) .
$$
Poisson brackets follow in the classical limit,
\begin{equation}
{\lim_{\hbar \rightarrow 0}} \fr{-i}{\hbar } \left[f(P,Q ) , g(P,Q ) \right]_\star
= \{ f(P,Q) , g(P,Q) \} \equiv
  \frac{\partial f}{\partial Q^I} \frac{\partial g}{\partial P_I}
-\frac{\partial f}{\partial P_I}\frac{\partial g}{\partial Q^I} .\label{ocl}
\end{equation}
For the  matrix observables  $M_{kl}(P,Q)$ and $N_{kl}(P,Q)$
the Moyal bracket can be defined as
\begin{equation}
\label{MM}
\left( \left[M(P,Q) , N(P,Q ) \right]_\star \right)_{kl}  =  M_{km}(P,Q ) \star N_{ml}(P,Q )
- N_{km}(P,Q ) \star M_{ml}(P,Q ) .
\end{equation}
In contrary to the observables which are functions, matrix observables in general are not Abelian,
so that the classical limit (\ref{ocl}) of their Moyal bracket (\ref{MM}) will yield, in general,
a term behaving as $\hbar^{-1}$ which would be singular.
However, if the observables are related to spin they may possess $\hbar$ dependence, then
the singularity which we mention is superfluous. To take into account this fact
 we define  the  ``semiclassical'' bracket
\begin{equation}
\label{cbr}
 \left\{M(P,Q ) , N(P,Q ) \right\}_C \equiv
- \frac{i}{\hbar}[M,N]+
\frac{1}{2} \{ M(P,Q ) , N(P,Q) \}
-\frac{1}{2} \{ N(P,Q ) , M(P,Q ) \} ,
\end{equation}
where one should retain the  terms
up to  $\hbar .$
The first term on the right hand side is the ordinary commutator of the matrices and the last two terms are Poisson brackets.
The bracket (\ref{cbr}) does not satisfy Jacobi identities. This is due to the fact that in its definition one keeps the first two terms of
the fully-fledged Moyal bracket (\ref{MM}). However, in our approach
the semiclassical limit  is taken after multiplying the observables by the star product (\ref{SCT}).
Hence, we should consider  the semiclassical limit of the Jacobi identity which is  the first two terms of the Moyal bracket relation
$$
- \frac{i}{\hbar}\left(\{M,\{ N,L \}_\star \}_\star +
\{N,\{ L,M \}_\star\}_\star  +
\{L,\{ M,N \}_\star\}_\star \right) = {\cal O}_{-1} \left(\frac{1}{\hbar}\right)+{\cal O}_0 (\hbar^0)+{\cal O}_1 (\hbar) +\cdots.
$$
In fact, one can show that the semiclassical limit of the Jacobi identity is satisfied
\beqar
{\cal O}_{-1} (\frac{1}{\hbar})+{\cal O}_0 (\hbar^0) & = &
-\fr{i}{\hbar}[M,[N,L]]+[M, \{ N , L \}]
-[M , \{ L , N\}]
+\{M ,[N,L]\}\nonumber \\
&&
 -\{[N,L],M\}
 + ({\rm cyclic\ permutations\ of}\  M,N,L) =0 .\nonumber
\eeqar
Nevertheless, once we perform quantization and deal with quantum operators choosing a realization of quantum phase space variables, we should impose that
they satisfy Jacobi identities.

The  first order matrix Lagrangian adequate to formulate spin dynamics in noncommutative coordinates  is
\beq
\label{lag}
L   = {\dot r}^\al \left[ \frac{p_\al}{2}\mathbb{I} +\rho A_\al (r) \right]
-\frac{{\dot p}^\al}{2} \mathbb{I}\left[r_\al  + \frac{\theta_{\al \beta}}{ \hbar} p_\beta  \right] - H_0(r,p)
\eeq
where
 $\al,\beta =1,\cdots ,{\rm d}. $
We would like to emphasize that $\A_\al $ is in general matrix valued. $\rho    $
denotes the related coupling constant and
$\mathbb{I}$ is the unit matrix.
 The constant, antisymmetric non-commutativity parameter $\theta_{\al \beta }$   appears divided by $\hbar$
to set  its dimension at $\left(\sf{length}\right)^2$.
The definition of canonical momenta
$$
\pi_r^\al =\frac{\del L}{\del \dot r_\al} ,\qquad \pi_p^\al =\frac{\del L}{\del \dot p_\al}
$$
yields the dynamical  constraints
\beqar
\psi^{1\al} & \equiv & \left(\pi_r^\al -\fr{1}{2}p^\al \right)\mathbb{I} -\rho \A^\al ,\lb{pcs1}\\
\psi^{2\al} & \equiv & \left(\pi_p^\al +\fr{1}{2}r^\al \right)\mathbb{I} +\frac{\theta_{\al \beta}}{ \hbar} p_\beta .\lb{pcs2}
\eeqar
By setting $P_I\equiv (\pi_r^\al , \pi_p^\al) , Q_I \equiv (r^\al , p^\al) $
in (\ref{cbr}) one can show that the constraints (\ref{pcs1}) and (\ref{pcs2})  obey the semiclassical brackets
\beqar
\{\psi^1_\al ,\psi^1_\beta \}_C & = & \rho F_{\al \beta} ,\nonumber \\
\{\psi^2_\al ,\psi^2_\beta \}_C & = & \frac{\theta_{\al \beta}}{ \hbar},\nonumber \\
\{\psi^1_\al ,\psi^2_\beta \}_C & = & -\delta_{\al \beta} .\nonumber
\eeqar
$\delta_{\al\beta}$ is  the Kronecker delta and $F_{\al \beta}$ is
the field strength,
\beq
F_{\al \beta} =
\fr{\partial A_\beta}{\partial r^\al}
-\fr{\partial A_\al}{\partial r^\beta}
-\fr{i\rho }{\hbar} [A_\al ,A_\beta ] ,\label{fst}
\eeq
where the last term is the ordinary matrix commutator. Thus, we may classify
$\psi^z_\al ;\ z=1,2,$ as second class constraints and the matrix whose elements are
\beq
\lb{cm}
{\cal C}^{zz^\prime}_{\al \beta}=
\{\psi^z_\al , \psi^{z^\prime}_\beta\}_C ,
\eeq
possesses the inverse ${\cal C}^{-1}$:
\beq
{\cal C}^{zz^{\prime \prime}}_{\al \gamma }
{\cal C}_{z\prime z^{\prime\prime}}^{-1\gamma \beta}=\delta_\al^\beta \delta_{z^\prime}^z .
\eeq
The inverse matrix elements  can be employed to define the ``semiclassical Dirac bracket" as
\begin{equation}
\label{sdb}
\{M,N\}_{CD} \equiv \{M,N\}_C -\{M,\psi^z\}_C \ {\cal C}^{-1}_{zz^\prime}\ \{\psi^{z^\prime},N\}_C  ,
\end{equation}
so that the constraints (\ref{pcs1}) and (\ref{pcs2})
effectively vanish.
The basic
classical relations between the phase space variables
following from (\ref{lag})
 can be established,
 at the first order in $\theta$ and keeping at most the second
order terms in $\rho ,$ as
\beqar
\{r^\al,r^\beta \}_{CD}  & = &
 \frac{ \theta^{\al\beta  }}{\hbar} , \lb{rr}\\
\{p^\al,p^\beta \}_{CD}  & = &
\rho  F^{\al\beta  }   -\frac{\rho^2}{\hbar}(F\te F)^{\al\beta} ,  \lb{yy}\\
\{r^\al,p^\beta \}_{CD}  & = &
\delta^{\al \beta } - \frac{\rho}{\hbar} (\te F)^{\al\beta}    \lb{ry}
\eeqar
where $(\te F)^{\al\beta} \equiv \te^{\al\ga}F_{\ga}^{\beta} ,
(\te F\te)^{\al\be} \equiv \te^{\al\ga} F_{\ga}^\si \te_{\si}^{\be}$.
We omitted the identity matrix $\mathbb{I}$ on the left hand sides.
Indeed, in the sequel we will not  write  $\mathbb{I}$ explicitly.

The brackets (\ref{rr})-(\ref{ry}) differ from the Poisson brackets up to commutators of matrices, so that for observables which are
not matrices they reduce to the ordinary Dirac brackets. Therefore, we can extend the canonical quantization rules to embrace the matrix observables
by substituting the basic brackets with the quantum commutators as
$\{~, ~\}_{CD} \rightarrow \frac{1}{i\hbar} [~ ,~]_q .$ To distinguish the matrix commutators and
quantum commutation relations we denoted the latter as $[~ ,~]_q .$
This yields the generalized algebra
\beqar
{[\hat r^\al ,\hat r^\beta ]_q}  & = &
 i \theta^{\al\beta  } ,\lb{rrq}\\
{[ \h p^\al,\h p^\beta ]_q } & = &
i\hbar \rho  F^{\al\beta  }   -i\rho^2(F\te F)^{\al\beta} , \lb{yyq}\\
{[\h r^\al,\h p^\beta ]_q } & = &
i\hbar \delta^{\al \beta } - i\rho (\te F)^{\al\beta} ,  \lb{ryq} \\
{[\h p^\al , \h r^\beta ]_q}  & = &
-i\hbar \delta^{\al \beta } + i\rho (F \te )^{\al\beta}   .\lb{yrq}
\eeqar
Note that, on the right hand side we keep the  first order $\theta$ contributions, so that  everything can only depend on
 $x_\al$,  defined as ${\h r}_\al|_{\theta =0} =x_\al .$
For Abelian gauge fields this type of algebra has already been considered in \cite{cnp} and a similar one in noncommutative space
for an electromagnetic field was discussed in \cite{dh} (see also \cite{dbg} and  references therein).

One can employ realizations of the algebra (\ref{rrq})-(\ref{yrq}) to introduce noncommutative coordinates.
To illustrate it, let us deal with the commutative case $\theta =0$ and let the gauge field be not a matrix but the 2--dimensional
electromagnetic one $ a_i = (-Br_2/2, Br_1/2),$
which leads to a constant  magnetic field transverse to the $(r_1,r_2)$-plane $B.$ For these choices the algebra becomes
\beq
\lb{rrqs}
{[\hat r_i ,\hat r_i ]_q}   = 0,\qquad
{[ \h p_i,\h p_j ]_q }  =
ie\hbar B \epsilon_{ij}, \qquad
{[\h r_i,\h p_j ]_q }  =
i\hbar \delta_{ij }  .
\eeq
A realization of the algebra (\ref{rrqs}) is
\begin{equation}
\label{kmom}
   {\h p}_i  =  -i\hbar \frac{\partial}{\partial r_i} +e a_i ,\qquad \h r_i = r_i .
\end{equation}
Through
the substitution of classical momenta with the realization (\ref{kmom}),
in the free Hamiltonian $H_0=p^2/2m ,$
the minimal coupling to the gauge field  in quantum mechanics  can be achieved  as
\begin{equation}
\label{hadmi}
H_{{\rm int}}\equiv H_0 (\h p , q)=\frac{1}{2m} \left(-i\hbar \frac{\partial}{\partial \vec r} +e\vec a \right)^2 .
\end{equation}
We will extend  this point of view to define quantum mechanics in noncommuting coordinates.

Although we will employ another realization   to propose
a Hamiltonian adequate to describe graphene  on the noncommutative plane,
let us present a  realization of (\ref{rrq})--(\ref{yrq}).
In terms of the covariant derivative
\beq
D_\al =-i\hbar \frac{\pa}{\pa x_\al} -\rho A_\al\equiv -i\hbar\nabla_\al  -\rho A_\al ,
\eeq
we can  realize the algebra (\ref{rrq})--(\ref{yrq}) by setting
\beqar
\h p_\al &=& D_\al - \fr{\rho}{2\hbar}F_{\al\beta}\te_{\beta\ga}D_\ga ,\lb{ps}\\
\h r_\al & =& x_\al -\fr{1}{2\hbar}\te_{\al\beta}D_\be  ,\lb{rs}
\eeqar
as far as $F_{\al\beta}$ are constant, commuting matrices which are equivalent to the conditions
\beq
\label{cons}
  -i\hbar\nabla_\al F_{\beta \ga}  -\rho [ A_\al , F_{\beta \ga}] = 0,\qquad  [F_{\al \beta },F_{\ga \delta }]= 0.
\eeq
 These conditions are also necessary to show that the realization (\ref{ps})--(\ref{rs})
satisfies the Jacobi identities.
We would like to emphasize that
this realization  is valid for either  Abelian or non-Abelian $A_\al .$
It
can be employed to introduce the related dynamical system in
noncommutative coordinates as
\beq
\label{pnc}
H(\te )\equiv H_{(0)}(\h r,\h p),
\eeq
where $H_{(0)}( r, p)$ is the free Hamiltonian appropriate to the considered system.
Indeed, this constitutes  an alternative method to the star-product approach of introducing
noncommutative coordinates  in quantum systems.

\section{Dirac particles in noncommutative space}

In graphene, around each  Dirac  point, which is the point at the corners of Brillouin zone,
the free Hamiltonian is
written as the massless Dirac-like Hamiltonian~\cite{semenoff,shon-ando}
\begin{equation}
\label{hadm}
H_{\sf D}^{(0)}(p , q)= v_{\sf F}\,   \vec p \cdot \vec\si
\end{equation}
for low energies and long wavelengths. Here,
$\vec p=(p_x,p_y)$ is the two-dimensional momentum operator and
$\vec \sigma=(\sigma_x,\sigma_y)$
where $\sigma_{x,y,z}$ are the  Pauli matrices acting on the states of two sublattices.
$v_{\sf F}$ is the Fermi velocity playing the role of the speed of light in vacuum.

We would like to deal with the dynamics of
the massless Dirac particle on
the noncommutative $(x,y)$-plane whose
free Hamiltonian is (\ref{hadm}) within the method presented in the previous section.
For this purpose let the gauge field be
\beq
\label{agt}
 A_i= -\frac{eB}{ 2} \epsilon_{ij} x_j + ik \epsilon_{ij}\si_j +l\si_i, \qquad i,j=1,2,
\eeq
which is non--Abelian.
The first term corresponds to the transversal, constant magnetic field $B$ and
the  others are spin-orbit-like coupling terms. However one should keep in mind that
for graphene $\vec \si$  act on  the states of sublattices, so that  though
$k$ and $l,$ respectively, like  the
coupling constants related to the Rashba and Dresslhauss spin-orbit interaction terms
for electrons, their effect is to give rise to terms proportional to $\sigma_z$ and unity
in the non--deformed Hamiltonian. In fact, shifting momenta in (\ref{hadm}) with the gauge field (\ref{agt})
yields the following  Dirac--like Hamiltonian
\beqar
H_{\sf D} & = & v_{\sf F}\,  (\vec p - \vec A) \cdot \vec\si \nonumber \\
&=& v_{\sf F}   \si_i \left(-i\hbar\nabla_i +\frac{eB}{ 2} \epsilon_{ij} x_j
\right) -2 v_{\sf F}(k\si_z +l). \label{hin}
\eeqar
We would like to get a $\theta$-deformation of this Hamiltonian employing the procedure outlined in the previous section.
Hence,
we set $\rho=1$ and by using the definition (\ref{fst}) we
obtain the field strength corresponding to  (\ref{agt}) as
\beq
\label{tfij}
F_{ij}=\left(eB+\frac{2 }{\hbar}(l^2-k^2)\sigma_z\right)\epsilon_{ij}.
\eeq
The algebra (\ref{rrq})--(\ref{ryq}) now becomes
\beqar
{[\hat r_i ,\hat r_j ]}  & = &  i \ep_{ij  } \theta , \lb{rrq1}\\
{[ \h p_i,\h p_j ] } & = & i\hbar \left(eB +\frac{2}{ \hbar}(l^2-k^2)\si_z\right)\ep_{ij}
+\left( ie^2B^2\te +\frac{4i}{ \hbar}eB\te (l^2-k^2)\si_z \right) \ep_{ij} ,\lb{yyq1}\\
{[\h p_i , \h r_j ]}  & = &
-i\hbar \delta_{ij } \left( 1 + \frac{\te}{ l_B^2} +(l^2-k^2)\frac{2\te}{ \hbar^2}\si_z\right) \lb{yrq1}
\eeqar
where $l^2_B=\frac{\hbar }{ eB} .$ We deal with small $l,k,$ so that we neglect the terms at the order $l^nk^m$ for $ n+m\geq 4 .$
Obviously, (\ref{agt}) and (\ref{tfij}) do not satisfy the conditions (\ref{cons}),
so that one cannot make use of the realization (\ref{ps}), (\ref{rs}).
Nevertheless, we accomplish a  realization of (\ref{rrq1})--(\ref{yrq1})  as follows:
\beqar
\hat p_i&=& \left[ 1 + \frac{\te}{ 2l_B^2} +(l^2-k^2)\frac{\te}{ \hbar^2}\si_z\right]
 \left(-i\hbar\nabla_i +\frac{eB}{ 2} \epsilon_{ij} x_j - ik \epsilon_{ij}\si_j -l\si_i\right)  \nonumber \\
&& +(l^2-k^2)\frac{2\te}{ \hbar^3}\epsilon_{nm} x_n (-i\hbar\nabla_m) (ik\epsilon_{ij} \si_j +l\si_i) ,\label{ar1} \\
\hat r_i&=& \left[ 1 + \frac{\te}{ 2l_B^2} +(l^2-k^2)\frac{\te}{ \hbar^2}\si_z\right] x_i
-\frac{\te}{ 2\hbar}\epsilon_{ij}\left(-i\hbar\nabla_j -\frac{eB}{ 2c} \epsilon_{jn} x_n\right)  \nonumber \\
&&- \frac{\te}{ \hbar^3} (l^2-k^2) \epsilon_{ij} \Big[ \left(ik \epsilon_{jn}\si_n +l\si_j\right) x_m^2
- 2  \left(ik \epsilon_{n m}\si_m +l\si_n\right) x_n x_j\Big] .  \label{ar2}
\eeqar
One can  demonstrate that (\ref{ar1}) and (\ref{ar2}) satisfy the Jacobi identities at the first order in
$\te$ and ignoring the terms at the order of
$l^nk^m$ for $ n+m\geq 4 .$

Through  the procedure outlined with (\ref{pnc}) $\theta$-deformation of the  Hamiltonian (\ref{hin}) can be achieved
by substituting momenta with the realization (\ref{ar1}) in the Hamiltonian (\ref{hadm}) as
\begin{equation}
\label{ncdham}
H_{\sf D}^{(\te)}= \frac{v_{\sf F} }{ 2} \left[ \h{\vec p} \cdot \vec\si +\left( \h{\vec p} \cdot \vec\si \right)^{\da} \right].
\end{equation}
Indeed, plugging (\ref{ar1})  into  (\ref{ncdham})  yields
\beqar
H_{\sf D}^{(\te)} &=& v_{\sf F} \left(1 + \frac{\te}{ 2l_B^2}\right) \left(-i\hbar\nabla_i +\frac{eB}{ 2} \epsilon_{ij} x_j
\right) \si_i  -2 v_{\sf F}\Big[ 1 + \frac{\te}{ 2l_B^2} +(l^2- k^2)\frac{\te}{ \hbar^2}\si_z  \nonumber \\
&& + 2(l^2- k^2) \frac{\te }{ \hbar^3}
\epsilon_{nm} x_n (-i\hbar\nabla_m)\Big]\, (k\si_z +l).  \label{gha}
\eeqar
One can observe that for $\te =0$, it  yields the Hamiltonian given in (\ref{hin}).
In \cite{gbmg} noncommutative structure emerging in graphene was studied where the noncommutativity parameter
considered is due to a lattice distortion term present in
the Hamiltonian. In fact, it is similar to the
mass term, so that it is related the our constant parameter $k.$
However, it is like a coordinate dependent mass term because  they also consider  a nonconstant distortion.

To establish energy eigenvalues it is convenient to write
(\ref{gha}) in terms of the complex variables $z=x+iy,\ \bar z=x-iy$ as
\begin{equation}
\label{hdc}
H_{\sf D}^{(\te)}
= \left( \begin{array}{c c}
g_+ - h_+ \frac{L_z}{ \hbar} & iK\left(-2\hbar\nabla_{z} + \frac{eB}{ 2}\bar z\right)\\
-iK\left(2\hbar\nabla_{\bar z}+ \frac{eB}{ 2} z\right) & g_- - h_- \frac{L_z }{ \hbar} \\
\end{array} \right)
\end{equation}
where  $L_z=-i\hbar\ep_{ij}x_i\nabla_j=\hbar (z\nabla_{z} -\bar z \nabla_{\bar z})$ is the angular momentum operator.
The involved constants are defined as
$$g_{\pm} = -2v_{\sf F}(l\pm k) \left[ 1+\frac{\te }{ 2l_B^2} \pm \frac{\te }{ \hbar^2}
(l^2 - k^2)\right] ,\qquad
h_{\pm} =4 v_{\sf F}\frac{\te }{ \hbar^2}(l\pm k) (l^2 - k^2) ,\qquad
 K = v_{\sf F}(1+ \frac{\te }{ 2l_B^2}).$$

To derive the eigenvalues  of  (\ref{hdc})  algebraically,
we introduce two pairs of annihilation and creation operators:
\beqar
&a=-\frac{il_B}{  \sqrt{ 2}\hbar }\left( 2\hbar\nabla_{\bar z}+\frac{eB}{ 2} z\right),\qquad
 a^\da = \frac{il_B}{  \sqrt{ 2}\hbar }\left( -2\hbar\nabla_{z} + \frac{eB}{ 2}\bar z\right) ,&\nonumber \\
&b= -\frac{il_B}{  \sqrt{ 2}\hbar }\left(2\hbar\nabla{ z}+ \frac{eB}{ 2} \bar z\right),\qquad
b^\da = \frac{il_B}{  \sqrt{ 2}\hbar }\left( -2\hbar\nabla_{\bar z} + \frac{eB}{ 2} z\right) ,& \nonumber
\eeqar
which  are mutually commuting and  satisfy the commutation relations
$$
[a, a^\da] = [b, b^\da] = 1.
$$
Hence, the Hamiltonian (\ref{hdc})
acquires the form
$$
H_{\sf D}^{(\te)}
= \left( \begin{array}{c c}
g_+ - h_+ (b^\da b- a^\da a) & \ti K a^{\da}\\
\ti K a & g_- - h_- (b^\da b- a^\da a)  \\
\end{array} \right)
$$
where
$\ti K =  2v_{\sf F}\hbar eB \left(1+ \frac{\te }{ 2l_B^2}\right).$
The eigenvalue equation
for the two component spinor
$$
H_{\sf D}^{(\te)} \left(
\begin{array}{c}
\psi_{1} \\
 \psi_{2}\end{array}\right)= E \left(
\begin{array}{c}
\psi_{1} \\
\psi_{2}\end{array}\right) .
$$
leads to two coupled equations
\beqar
\left[g_+ - h_+ (b^\da b- a^\da a) -E\right]\psi_{1} & = & - \ti K a^\da \psi_{2} ,\label{e1}\\
\left[g_- - h_- (b^\da b- a^\da a) -E\right]\psi_{2} & = &- \ti K a \psi_{1}. \label{e2}
\eeqar
After some calculation, one can show that the  equation satisfied by  the spinor component $\psi_1$
takes the form
\beq
\label{sorder}
 \left[E^2  +4E\left( Kl + \frac{\te v_{\sf F}}{ \hbar^2} l(l^2-k^2) (1+  2b^\da b+ 2a^\da a)   \right)
-{\ti K}^2  a^\da a + 4 K^2(l^2-k^2) \right] \psi_{1} =0.
\eeq
To draw the energy eigenvalues, let us write the  state corresponding to the
spinor component $\psi_1$  as
\beq
\label{hos}
\mid \psi_1 \rangle =\mid n,m\rangle = \frac{1}{\sqrt{n! (m+n)!}}(b^\da)^{m+n} (a^\da)^n \mid 0\rangle ;
\eeq
$n,m=0,1,2\cdots$, and by definition  $a |0> = b | 0 > =0.$
In the complex plane (\ref{hos}) yields
$$
\langle z,\bar z\mid n,m\rangle  =
 N_{mn} z^{m}L^{m}_{n}\left(\frac{z\bar{z}}{2}\right)e^{-\frac{1}{4}z\bar{z}}
$$
where  $L^{m}_{n}$ are the Laguerre polynomials
and $N_{mn}$ are the normalization constants whose explicit forms are not needed.

Obviously, (\ref{hos}) satisfies the relations
\beqar
&& (b^\da b- a^\da a)\mid n,m\rangle = m \mid n,m\rangle ,\nonumber\\
 && a^\da a \mid n,m\rangle = n \mid n,m\rangle  ,\nonumber
\eeqar
where $m$ and $n$ are the quantum numbers corresponding, respectively, to the angular momentum eigenvalues
and the Landau levels.
Now,
(\ref{sorder}) can be solved to deduce the  energy spectrum
as
\beq
E_{n,m}(k,l,\te, B) = \pm 2v_{\sf F}\left(1+ \frac{\te }{ 2l_B^2}\right) \sqrt{\frac{\hbar^2 }{ 2l_B^2}n +k^2}
 -2v_{\sf F} l\left[1+ \frac{\te }{ 2l_B^2}
 + \frac{\te}{ \hbar^2} (l^2-k^2) (2m+1) \right] . \label{eei}
\eeq
Moreover, one can show that the corresponding spinor components are given by
$$
    \Psi_{n,m}=\left(
\begin{array}{c}
   \mid n,m\rangle\\
s'  \mid n-1,m+1\rangle\\
\end{array}
\right)
$$
with the convention $\psi_{-1,m}\equiv 0.$
Here
 $s'$ is a constant, which can be read from (\ref{e2}).

\section{ Shubnikov-de Haas effect}

The Shubnikov-de Haas (SdH) effect is a magnetotransport phenomena that occurs in materials in
a strong magnetic field of about $1\ {\rm Tesla}$ and for low temperature about few kelvins~\cite{haas}.  It
is an oscillatory dependence of the electrical resistivity  of a metal or a semiconductor as a function
of  the applied constant magnetic field.
More precisely, the SdH effect is produced by the oscillations of the density of states
at the Fermi level.
The mechanism can be understood for metals considering Landau levels~\cite{kittel} which are the
energy levels of electrons in the presence of a magnetic field. If the electrons fill the energy levels up to  the level $n+1,$
the Fermi energy which is equal to the chemical potential at absolute zero,
will lie in this level. As the magnetic field increases the degeneracy of the  Landau level
increases. Thus the electrons move to the level $n,$ depopulating the level $n+1,$ so that the Fermi energy is decreased.
Now, increasing magnetic field leads to less populated Fermi level until all electrons migrate to the lower energy level.
Hence, the conductance or the resistivity
will oscillate as a function of the external magnetic field.
The maxima of the SdH effect
occur  at the magnetic fields $B_N$
which can be calculated by equating
the energy level corresponding to the index $N$  with
the chemical potential $\mu$ (Fermi energy). Hence,
the relation between $N$ and $B_N$ predicted by our approach
is established as
\beq
\label{nfield}
N = \frac{1}{ 2e\hbar B_N} \left[\frac{\mu^2}{ v_{\sf F}^2} + 4\frac{\mu }{ v_{\sf F} } l + 4(l^2-k^2) \right] +\frac{\ti{\te}}{ \hbar^2}
\left[\frac{ 2}{ e\hbar B_N}l
 (k^2-l^2) (2m+1) +\frac{\mu}{ 2 v_{\sf F}}  +l \right] ,
\eeq
by solving the  equation  $E_{N,m}(k,l,\te , B_N)= \mu$ obtained from (\ref{eei}).
For convenience,  we rescaled the noncommutativity
parameter as
$$
\theta=-\frac{v_{\sf F}}{\mu}\ti{\te}.
$$

To analyze the  SdH effect in graphene within our formulation we shall choose
the involved parameters  adequately.
To start with, we require that the spin-orbit-like coupling constants obey
$$l= -\frac{\mu }{ 2 v_{\sf F}} + k.$$
With this choice   (\ref{nfield}) is simplified and takes the form
\beq
\label{nsp}
N = \frac{\tilde\te}{ \hbar^2 }    \left(\frac{B(m,k) }{  B_N}  +k \right)
\eeq
where we defined
$$
B(m,k)= \frac{\mu  }{ v_{\sf F} e\hbar }
 \left(\frac{\mu }{ 2v_{\sf F}}-2k\right) \left(\frac{\mu }{ 2v_{\sf F}} -k\right) \left(1+2m \right) .
$$

Since the noncommutativity parameter $ \te$ is a free parameter, it can be
fixed in diverse fashions.
However, one should keep in mind that its value should be consistent with the approximation
of retaining the terms up to the first order in $\te .$
In particular, for the limiting values of $B_N$, we propose to choose
 $\ti\te$  as
\beq
\label{asp}
\ti\te(B)= \left\lbrace
\begin{array}{ll}
{\beta / B_N}, & \qquad B_N>B(m,k) /k \\
\gamma B_N,  & \qquad B_N\ll B(m,k) /k
\end{array} \right\rbrace
\eeq
where $\gamma, \beta $ are two constants and we assume that $k\neq 0.$
We can  analyze (\ref{nsp}) separately for each case given in (\ref{asp}).
For $B_N >B(m,k) /k$ we deduce  the  behavior
\beq
\label{nu}
N_> = \frac{\be k}{ \hbar^2}\frac{1}{ B_N},
\eeq
by neglecting a term behaving as $1/ B_N^2.$
Thus, for large  $B_N,$
 $N$ changes linearly with respect to $1/ B_N$.
However, in the second case, $B_N\ll B(m,k) /k,$
 $N$ leads to the constant value
\beq
\label{nd}
N_< = \frac{\ga B(m,k) }{ \hbar^2 } .
 \eeq

Let us link these considerations to the  experimental observations of
\cite{berger}.
They obtained the limiting values
\beq
\label{nex}
N_{\rm exp} = \left\lbrace
\begin{array}{ll}
{B_0 / B_N}, & \qquad B_N> 2.5 \ {\rm T} \\
25 , & \qquad B_N\ll 2.5 \ {\rm T}
\end{array} \right\rbrace
\eeq
where the constant is given by
$$
B_0 =\frac{\mu^2}{2e\hbar v_{\sf F}^2}\approx 35 \ {\rm T} .
$$
This fixes the ratio
$$
\frac{\mu }{ v_{\sf F}}  \approx 34 \times 10^{-27} \ {\rm kg.m/s}.
$$

Now, we would like to determine the value of the noncommutativity parameter $\te $
comparing (\ref{nex}) with (\ref{nu}) and (\ref{nd})  for $m=0.$ The other values of
 $m$ can be treated similarly. First of all
observe that we may impose
\beq
\label{fbk}
\frac{B(0,k)}{k}=2.5 \ {\rm T} .
\eeq
To simplify let  $k=(\mu /2v_{\sf F}) \delta $,  so that (\ref{fbk}) yields
the equation
$$
2 \delta^2 -\left(3+\frac{2.5}{ B_0} \right) \delta +1 =0
$$
whose solutions are
$$
\delta \approx 0.77\pm 0.3.
$$
Hence, we may set
$$
k = 1 \times 10^{-26} \ {\rm kg.m/s}
$$
which implies to choose
$$
\beta \approx 4 \times 10^{-41} {\rm JmsT},\qquad
\gamma \approx 1 \times 10^{-41} {\rm JmsT^{-1}}.
$$
It worths to observe that the magnitude of the noncommutativity parameter for the limiting cases (\ref{asp}) reads
$$
|\te (B)|= \left\lbrace
\begin{array}{ll}
B_N^{-1} \times 10^{-15} \ {\rm m^2},\ \ &  B_N> 2.5 \ {\rm T}  \\
B_N \times 10^{-16} \ {\rm m^2},  \ \ &  B_N\ll 2.5 \ {\rm T}
\end{array} \right\rbrace  .
$$
Therefore, there is no conflict with keeping the terms up to the first order in $\theta .$
Until now we dealt with the  values of $\theta$ for the limiting values of the magnetic field $B_N.$
However, we can also choose it appropriately
for all values of $B_N.$ Inserting
the choice (\ref{fbk}) into (\ref{nsp}) yields
\beq
\label{nsp1}
N = \frac{\tilde\te k}{ \hbar^2 }    \left(\frac{2.5 \ {\rm T} }{  B_N}  +1 \right) .
\eeq
To write the full expression for $\ti \te$, let us
introduce the Heaviside step function
$$
H(x)= \left\lbrace
\begin{array}{ll}
0, \qquad & x<0 \\
 1/2,\qquad &  x=0 \\
 1, \qquad & x>0
\end{array} \right\rbrace
$$
which can be given analytically as~\cite{hev}
$$
H(x)=\lim_{t\rightarrow 0}\left[\frac{1}{2} +\frac{1}{ \pi} \tan^{-1}\frac{x}{ t}\right] .
$$
We choose the noncommutativity parameter to be
\beqar
\ti \te &=&\frac{\hbar^2/ k}{1+2.5 B^{-1}_N}\Bigg\{ 35  B^{-1}_N \left[
\frac{1}{2} +\frac{1}{\pi} \tan^{-1}\left(\frac{0.4 - B^{-1}_N}{ 0.01 }\right) \right]  +\frac{53 }{ 0.9 + B_N}
\left[\frac{1}{2} +\frac{1}{ \pi} \tan^{-1}\left(\frac{ B^{-1}_N-0.4}{ 0.01 }\right)\right] \nonumber \\ &&
\times \left[\frac{1}{2} +\frac{1}{ \pi} \tan^{-1}\left(\frac{0.83 - B^{-1}_N}{ 0.01}\right)\right]
+24.8
\left[\frac{1}{2} +\frac{1}{ \pi} \tan^{-1}\left(\frac{B^{-1}_N-0.83}{ 0.01}\right)\right] \Bigg\} \label{tef}
\eeqar
which produces the limiting values correctly
and with this choice (\ref{nsp1}) yields  Figure 1 (using MATHEMATICA).
Indeed, we have chosen (\ref{tef}) appropriately so that the Landau plot
of the peaks given in Figure 1 matches  well  with the experimental one obtained in \cite{berger}.
Moreover,
one can check that the order of magnitude of
the noncommutativity parameter is $\theta \approx 10^{-16}\ {\rm m^2},$ so that it is in accord with the
approximation of ignoring the second order terms in $\te .$

\begin{figure}
\begin{center}
\includegraphics{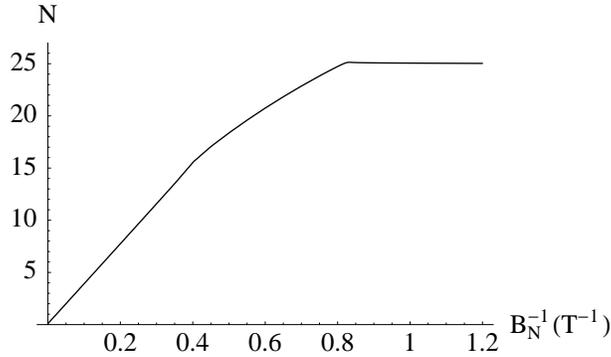}
\end{center}
\caption{Landau plot of the maxima of SdH oscillations.}
\label{f.lbl}
\end{figure}

\section{Discussions}
The results which we obtained are twofold:
\begin{itemize}
 \item
 A new method of introducing noncommutative coordinates into quantum mechanics is established.
\item An analytic method of obtaining confinement of massless Dirac particles in graphene is proposed.
\end{itemize}

We  introduced a generalized  algebra of quantum phase space operators in noncommuting space on
general grounds with momenta involving non-Abelian gauge fields. This constitutes an alternative
to the custom method of introducing noncommuting coordinates by star products. It may lead to some new features of
quantum mechanics in noncommutative coordinates. Moreover, it should be possible to extend it to field theory formulations.
These are currently under inspection.

We considered a two-dimensional  space by a particular choice of   gauge
fields. A realization of  the associated  algebra is presented
and employed to obtain  a massless Dirac-like Hamiltonian on the noncommutative plane.
Its energy eigenvalues are
established. Through an appropriate  choice of the
noncommutativity parameter $\te $ we showed that this energy spectrum is  adequate to accomplish
the experimentally observed behavior of the SdH oscillations in graphene,
which are known to result due to the confinement of its charge carriers which are  massless Dirac particles.
Obviously, our main objective is to employ this noncommutative theory to understand those features of graphene
which are not well understood within other formalisms.
This work should be considered as the first step in this direction. We obtained a satisfactory noncommutative
version of Dirac-like theory of graphene which led to some predictions. One of the next steps would be to obtain
a field theory in terms of the Hamiltonian (\ref{gha}), which can be used to introduce other interactions like the
spin of electron as in \cite{km} into
the noncommutative theory.

\section*{Acknowledgments}
We are grateful to Professor K. R. Sreenivasan and Professor S. Randjbar--Daemi for the  hospitality
at the Abdus Salam ICTP, Trieste--Italy, where the main part of this work
has been done.


\end{document}